\documentclass[9pt,twocolumn,twoside]{gsajnl}
\usepackage{epstopdf}
\usepackage{caption}
\usepackage{threeparttable}
\usepackage{cleveref}
\usepackage{subcaption}
\usepackage{graphics}
\usepackage{orcidlink}
\usepackage{tikz} 
\usetikzlibrary{matrix}
\usepackage{breqn}
\makeatother
\restylefloat{table}
\articletype{inv}
\usepackage{amsmath}      
\usepackage{graphicx}
\newlength\imagewidth
\newlength\imagescale
\usepackage{multirow, booktabs}
\usepackage{booktabs}
\usepackage{longtable,needspace}
\usepackage{float}
\usepackage{listings}
\usepackage{booktabs}
\usetikzlibrary{positioning}

\runningtitle{Optmization of breeding programs} 
\runningauthor{Hassanpour \textit{et al.}}
\title{Optimization of breeding program design through stochastic simulation with evolutionary algorithms}
\author[1,2,$\ast$]{Azadeh Hassanpour \orcidlink{0000-0003-1976-3457}}
\author[1,2,3]{Johannes Geibel \orcidlink{0000-0001-7172-3263}}
\author[1,2]{Henner Simianer \orcidlink{0000-0002-7551-3797}}
\author[4]{Antje Rohde \orcidlink{0000-0002-1574-1843}}
\author[1,2,5]{Torsten Pook \orcidlink{0000-0001-7874-8500}}

\affil[1]{University of Goettingen, Department of Animal Sciences, Animal Breeding and Genetics Group, Albrecht-Thaer-Weg 3, 37075, Goettingen, Germany}
\affil[2]{University of Goettingen, Center for Integrated Breeding Research, Carl-Sprengel-Weg 1, 37075, Goettingen, Germany}
\affil[3]{Friedrich-Loeffler-Institut, Institute of Farm Animal Genetics, 31535 Neustadt, Germany}
\affil[4]{BASF Belgium, Coordination Center – Innovation Center Gent, 9052 Ghent, Belgium
}
\affil[5]{Wageningen University \& Research, Animal Breeding and Genomics, P.O. Box 338, 6700 AH Wageningen, Netherlands}

\correspondingauthoraffiliation[$\ast$]{azadeh.hassanpour@uni-goettingen.de}

\begin{abstract}
The effective planning and allocation of resources in modern breeding programs is a complex task. Breeding program design and operational management have a major impact on the success of a breeding program and changing parameters such as the number of selected /phenotyped /genotyped individuals in the breeding program will impact genetic gain, genetic diversity, and costs. As a result, careful assessment and balancing of design parameters is crucial, taking into account the trade-offs between different breeding goals and associated costs. In a previous study, we optimized the resource allocation strategy in a dairy cattle breeding scheme via the combination of stochastic simulations and kernel regression, aiming to maximize a target function containing genetic gain and the inbreeding rate under a given budget. However, the high number of simulations required when using the proposed kernel regression method to optimize a breeding program with many parameters weakens the effectiveness of such a method. In this work, we are proposing an optimization framework that builds on the concepts of kernel regression but additionally makes use of an evolutionary algorithm to allow for a more effective and general optimization. The key idea is to consider a set of potential parameterizations of the breeding program, evaluate their performance based on stochastic simulations, and use these outputs to derive new parametrization to test in an iterative procedure. The evolutionary algorithm was implemented in a Snakemake pipeline to allow for efficient scaling on large distributed computing platforms. The algorithm achieved convergence to the same optimum with a massively reduced number of simulations. Thereby, the incorporation of class variables and accounting for a higher number of parameters in the optimization pipeline leads to substantially reduced computing time and better scaling for the desired optimization of a breeding program.
 
\end{abstract}


\dates{\rec{xx xx, 2024} \acc{xx xx, 2024}}

\usepackage{titlesec}
\titleformat{\section}%
  [hang]
  {\normalfont\bfseries\Large}
  {}
  {0pt}
  {}
\renewcommand{\thesection}{}
\renewcommand{\thesubsection}{\arabic{subsection}}

\begin{document}
\maketitle
\thispagestyle{firststyle}
\vspace{-13pt}

\section*{Introduction}
\lettrine[lines=2]{\color{color2}W}{}ith the rise of genomics, advancements in biotechnology, statistical modeling, and shifts in market demands, breeding programs have undergone a substantial transformation in recent decades. The strategic combination of these advancements empowers breeders to refine and adapt their breeding strategies. As a result, modern breeding programs can be considered as a complex resource allocation problem to manage both short-term genetic gain and long-term sustainability \citep{Henryon.2014, Berry.2015, Hickey.2017, Simianer.2021}. Breeding programs require substantial investments in both resources and time. Moreover, the consequences of breeding decisions may only be evident after several years. In addition to operating costs and complex logistics associated with it, breeders must consider a variety of design parameters related to specific breeding objectives, taking into account the potential risks and uncertainties associated with each decision and weighing the costs and benefits of each possible resource allocation \citep{jannink2023insight}. 
\par
This problem is further complicated by the fact that breeding actions or changes to breeding program parameters are highly interdependent, and a change in one step of the breeding program will impact a multitude of key characteristics (e.g., genetic gain, genetic diversity, cost) of the breeding program \citep{Simianer.2021}. Therefore, breeders seek to understand and assess the uncertainties that are inherent in predicting the outcomes of their breeding programs, given the long-term nature and complexity of the breeding process \citep{harris1984animal, Mi.2014, Simianer.2021}. A strategy that recently gained in popularity is to use stochastic simulation to assess breeding program design before practically implementing them \citep{Henryon.2014, CovarrubiasPazaran.2021}. For this, a variety of software can be used, including MoBPS \citep{Pook.2020}, AlphaSim \citep{Faux.2016, Gaynor2021}, Adam \citep{Liu.2018}, and QMsim \citep{Sargolzaei.2009}. 
\par
However, the optimization of a breeding program design using stochastic simulation is complicated by the fact that the output of a simulation is only the realization of a stochastic process. Thus, multiple replicates are necessary to reliably estimate the expected outcomes of a breeding scheme \citep{Bancic2023}. Since breeding programs often involve numerous parameters, it is not feasible to simulate all possible breeding designs many times, as simulating a real-world breeding scheme can be computationally expensive. Therefore, analysis of breeding program designs using stochastic simulations is usually limited to a couple of potentially interesting scenarios and research studies focusing on very specific aspects of breeding design \citep{lorenz2013resource, Henryon.2014,Hickey.2014,Woolliams.2015,Gorjanc.2018,Moeinizade.2019,Wellmann.2019,Allier.2020, Buttgen.2020,  Duenk.2021, OjedaMarin.2021, Moeinizade.2022}. 
\par

Recently, we introduced a framework for optimizing breeding program designs to address and generalize different aspects of the breeding program more effectively \citep{hassanpour2023optimization}. In this study, the emphasis was on providing a general optimization framework that facilitates the simultaneous optimization of multiple design parameters within breeding schemes. In the process of optimizing a breeding program, one begins with a random search algorithm to explore disparate areas of the search space to examine a broad range of parameter values and obtain a preliminary set of different breeding programs for optimization. 
\par
Subsequently, kernel regression is employed by fitting a local regression curve to the data points (simulations), which are essentially a weighted average in which more similar breeding schemes are weighted stronger. Kernel regression smooths the results derived from the initial stage to filter out the noise and create a more discernible understanding of the potential regions where optimal solutions may be found. Following the application of kernel regression, the smoothed data provides an indication of the prospective 'optima' within the search space. The search is then concentrated on these narrowed-down regions, resulting in a substantial reduction of the overall search space. Finally, this entire process is performed iteratively with each successive iteration using the refined results of the previous round.
\par
While this approach has proven effective in improving optimization results, its application is constrained to optimizing only a limited number of parameters. As the number of parameters for optimization increases, the computational demands for performing a sufficient number of simulations to obtain a broad coverage of the search space increases exponentially \citep{Hardle.1997, gutjahr2016}. Additionally, we are faced with the challenge of a procedure that requires manually narrowing down the search space through iterative steps and visual inspection. Recognizing these challenges, there is a need to create a new optimization framework that requires fewer simulations and is fully automated. 
\par 
Traditional optimization techniques such as the steepest descent method, conjugate gradient method, and quasi-Newton method \citep{Kiefer.1952, back2008evolutionary, burke2020gradient} tend to struggle in the settings of breeding planning with high dimensional search spaces and stochasticity in the evaluation of an objective function. Stochastic optimization techniques \citep{Fouskakis2007} provide a framework to cope with exactly these challenges and include techniques such as genetic and evolutionary algorithm \citep{pierreval1997using, alberto2002optimization}, simulated annealing \citep{ahmed1997optimizing}, and Bayesian optimization \citep{Schonlau.1998}.
\par
Previous research in the field of breeding planning is however limited to the application of Bayesian optimization in simplified settings with a fixed budget and only use of continuous variables \citep{diot2023bayesian, jannink2023insight}. A further issue for the optimization is that the evaluation of the objective function is computationally very expensive. Therefore the chosen optimization technique should allow several evaluations to be carried out in parallel. 
\par
When optimizing complex problems with a high number of parameters and large search spaces, evolutionary algorithms (EAs) have gained popularity due to their ability to address the mathematical complexities inherent to real-world optimization problems, i.e., mixed class and continuous decisions, multiple objectives, uncertainty, computationally demanding simulations, etc. \citep{Holland.1992, Back.2000, Deb.2001, Michalewicz.2004, Sivanandam.2008, Eiben.2015, Katoch.2021, jeavons2022design}.
\par
The fundamental idea behind an EA is very similar to breeding programs with potential parametrizations in the pipeline corresponding to the individuals/breeding nucleus in a breeding program. Evaluating these parameter settings on the objective function subsequently corresponds to phenotyping. Following this evaluation, the algorithm generates new potential parameter settings to be tested in the subsequent iteration (generation). In the simplest case, this can be a combination of parameter settings that were previously evaluated as the most promising (recombination) and minor modification of settings (mutation). Yet, many existing EAs are often problem-specific \citep{Slowik.2020}, and depending on the nature, complexity, and dimensionality of the problem to be solved, different variations of recombination/mutation operators are used for optimization \citep{Sipper.2018}.
\par
Considering the limitations and challenges mentioned above, the objective of this study is to introduce a new EA framework designed to accommodate the complexity and diversity of breeding programs with varying inputs, capable of optimizing breeding programs with both continuous and discrete decision variables along the example of a dairy cattle breeding scheme suggested in \citet{hassanpour2023optimization}. Our proposed framework is adaptable to any breeding program, regardless of the species, methodology, resources, or genetic traits involved. We provide herewith a new tool for a variety of breeding objectives and therefore applies to any plant or animal breeding program as long as it can be simulated/evaluated via stochastic simulations.

\section*{Materials and methods}
\begin{figure*}
  \centering
  \includegraphics[width=500pt, height=300pt]{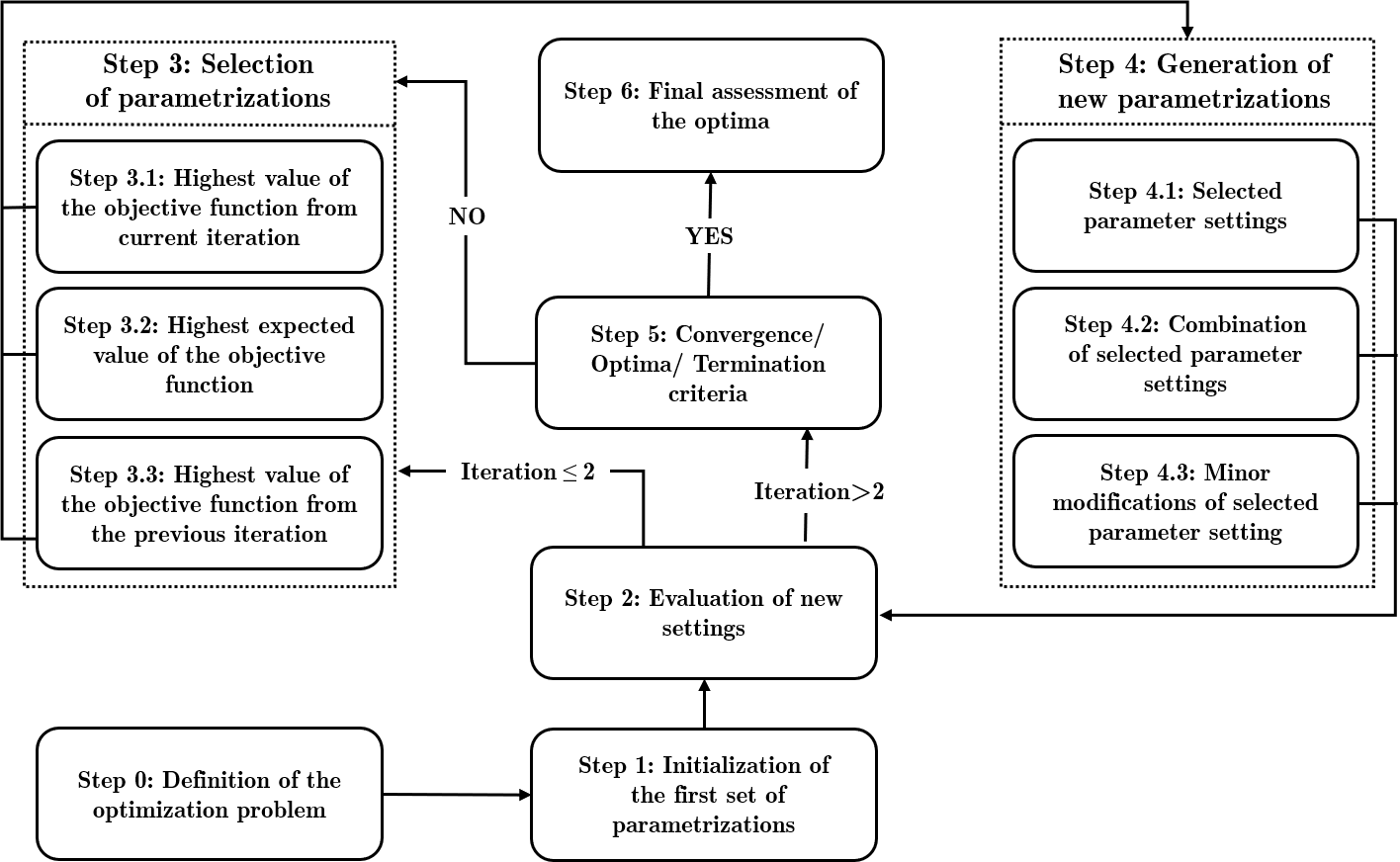}
  \caption{Procedure proposed for optimization via evolutionary algorithm.}  
  \label{fig:scheme}
\end{figure*}

In this study, we introduce a comprehensive pipeline for optimizing breeding scheme designs using an EA. The EA is structured as an iterative process, wherein Steps 2 to 5 are reiterated until a termination criterion is met. For illustrative purposes, we will outline the individual steps of the algorithm using the same dairy cattle breeding program previously examined in \citet{hassanpour2023optimization}. Figure \ref{fig:scheme} provides a schematic summary of the overall pipeline of our EA framework, representing key steps and their interconnections within the optimization process. 

\subsection*{Step 0: Definition of the optimization problem}~\\
Firstly, the breeding problem is formulated as an optimization problem. This involves defining the breeding objectives, practical constraints, and decision variables that govern the breeding strategy.
\par
In the following, we consider two main types of decision variables. Firstly, variables with a limited number of possible realizations that can be viewed as categorical features of a breeding program. An example of this is the question of whether genomic selection or marker-assisted selection is applied in a specific step of the breeding program. Secondly, we are considering variables that can take values from a continuous scale, or at least a large number of discrete realizations. This may include the number of candidates selected, phenotyped, and genotyped, or weights in a selection index. For the sake of simplicity, we will subsequently refer to these variables as continuous variables.

\subsubsection*{Scenario 1 - Traditional dairy cattle scheme}
\begin{figure*}
  \centering
  \includegraphics[width=270pt, height=170pt]{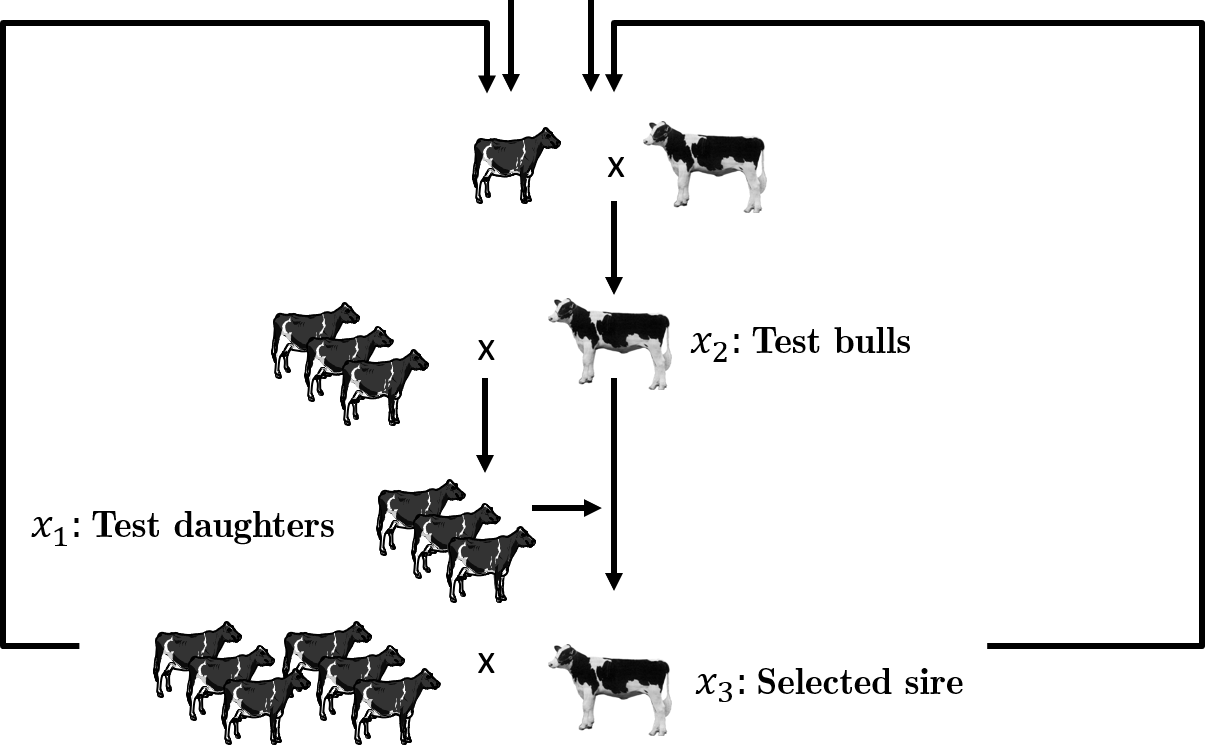}
  \caption{A dairy cattle breeding scheme.}  
  \label{fig:cow}
\end{figure*}

~\\
Here, we are considering a traditional dairy cattle breeding scheme, illustrated schematically in Figure \ref{fig:cow}. The three parameters we are here considering for optimization are:

\begin{enumerate}
\item $x_1$: number of test daughters
\item $x_2:$ number of test bulls
\item $x_3$: number of selected sires
\end{enumerate}

For simplification purposes, we are explicitly not considering the genotyping as commonly done in dairy cattle breeding in the last 15 years \citep{Schaeffer.2006} and are only considering a single quantitative trait (milk yield, with heritability ($h^2$) of 0.3). 

As constraints, the breeding program at hand is limited by an annual budget of 10,000,000 Euros with housing costs of 3000 Euros per bull and 4000 Euros per cow. To avoid unnecessary computations, unreasonable settings like the use of too many bulls or selecting just a single bull per year are excluded by additional constraints:

\begin{linenomath}
\begin{gather*} 
 x_1 + x_2 + x_3 \geq 0\\
 100 \leq x_2 \leq 700\\
 3 \leq x_3  \leq 30\\
 4000x_1 + 3000x_2 - 10000000 \leq 0 \quad\
\end{gather*}
\end{linenomath}

Exactly as in our previous study \citep{hassanpour2023optimization}, the objective function ($m$) is a linear combination of the expected genetic gain ($g$) and the expected inbreeding level ($f$) after 10 generations (5 years of burn-in + 10 years of future breeding), to prioritize/weigh between the genetic gain and diversity:

\begin{linenomath}
\begin{dmath*}
m(x) = g(x) - 50 \times f(x)
\end{dmath*}
\end{linenomath}

Subsequently, we showcase three additional examples to highlight the versatility of our EA framework, each representing a modification of the baseline (Scenario 1). These alternative scenarios are detailed towards the end of this section. 

\subsection*{Step 1: Initialization of the first set of parametrizations}~\\
To initialize the evolutionary pipeline it is necessary to generate a starting population of potential breeding program designs to consider. For this, we generally propose to use a generalized Bernoulli distribution for class variables and a uniform distribution for continuous variables. It should however be noted that depending on constraints a more sophisticated sampling procedure might be required, especially when the variables are interdependent, and logical dependencies for each decision may come from other decisions that share the same resources. 
\par
For example, let's consider a scenario related to housing capacity. Suppose there is a maximum stable capacity of 3000 individuals. Among these, there can be between 500 and 1000 males and 500 to 2500 females. Now, if we decide to use 1000 males during the sampling process, a logical constraint emerges, where the number of females cannot exceed 2000 to fulfill the overall capacity of 3000. Thus, it is important to follow these logical rules during the first step of sampling and throughout the process of optimization.
\par
For our toy example, an initial set of 600 parameterizations is generated with an additional scaling step on variables $x_1$ and $x_2$ to ensure that the budget constraint is met. For instance, if $x_1$ = 1025, $x_2$ = 300, and the total cost of the breeding program is 5,000,000 Euros. Given the budget is 10,000,000, with no advantages of underspending, both values are doubled. In case our budget constraint can not be exactly met due to rounding, a breeding program with costs slightly below the maximum cost is used, by first rounding down and then marginally increasing parameters as much as possible.

\subsection*{Step 2: Evaluation of new settings}~\\
In the next step, the suitability of each parametrization, meaning each breeding program design, needs to be evaluated regarding the breeding objective as given by the objective function $m$. For this evaluation, each respective breeding program is simulated via stochastic simulation. We here use the R package MoBPS \citep{Pook.2020} with scripts given in Supplementary File S1, but other simulation tools can be integrated seamlessly by the user. 

\subsection*{Step 3: Selection of parametrizations}~\\
Based on the results from the previous step, we want to identify the most promising areas of the search space to investigate further in later steps. For this, we select previously simulated parameter settings that will be used as parents of the next iteration, considering three strategies. The number of selected parents based on each step varies based on the iterations of the EA with values given below for iteration 11 onwards (see Table \ref{tab:parameter_values2}). In these iterations, 30 out of 300 parameterizations are selected.

\begin{table*}[h]
  \centering
\begin{threeparttable}
    \caption{Parameter setup for the number of selected and generated settings}
    \label{tab:parameter_values2}
    \begin{tabular}{cccccccc}
      \hline
      Iteration & \multicolumn{3}{c}{Selected settings} & \multicolumn{3}{c}{Generated new settings} \\
      \cmidrule(lr){2-4} \cmidrule(lr){5-7}
      &  Step 3.1 & Step 3.2 & Step 3.3 & Step 4.1 & Step 4.2 & Step 4.3\\
      \hline
      2-3  & 70 & 30 & 0 & 100 & 200 & 0\\
      4-10 & 30 & 15 & 5 & 50 & 170 & 80\\
      11-40 & 20 & 7 & 3 & 30 & 180 & 90\\
      \hline
    \end{tabular}
  \end{threeparttable}
      \begin{flushleft}
      Note: Step 3.1 presents individual simulations with the highest objective function value from the current iteration, Step 3.2 presents individual simulations with the highest expected value of the objective function determined by kernel regression, and Step 3.3 presents individual simulations with the highest objective function value from the previous iteration and previous optima. Step 4.1 represents the sum of Steps 3.1, 3.2, and 3.3. Step 4.2 represents the number of new settings (simulations) generated through a combination of selected parameterizations. Step 4.3 represents the number of new parameterizations created through minor modifications of selected settings. 
    \end{flushleft}
\end{table*}

\paragraph*{\normalsize \textit{Step 3.1: Highest value of the objective function}}~\\
20 of the 30 parental parametrizations are selected based on the value of the objective function that was derived solely based on the simulation of the parameterization itself. This guarantees that the best-performing parameterizations are prioritized in the reproduction process, subsequently enhancing the likelihood of generating successful parametrizations.

\paragraph*{\normalsize \textit{Step 3.2: Highest expected value of the objective function}}~\\
Secondly, we select seven candidates based on the highest expected value of the objective function. For this, we are employing the kernel regression method suggested in \cite{hassanpour2023optimization}. Such selection gives the advantage of instead of evaluating each candidate individually, kernel regression computes a weighted average of performance values for multiple candidates and in contrast to Step 3.1 will not be biased toward regions with more parameterizations tested overall. In contrast to \citet{hassanpour2023optimization}, we are here proposing the use of an adaptive bandwidth by using the empirical standard deviation in the individual parameter in the current iteration. 

\paragraph*{\normalsize \textit{Step 3.3: Highest value of the objective function from the previous iteration}}~\\
Finally, three parameterizations from previous iterations are used. This typically refers to the best-performing candidates from the last iteration, however, if the expected performance of any previously suggested optima 
 from any iteration based on kernel regression (see Step 5) is higher than all parametrizations of the current iteration, these are added instead. Hereby, the risk of discarding potentially valuable solutions and high-performing candidates due to stochasticity is reduced. 

\subsubsection*{Diversity management}~\\
Similar to real-world breeding programs, we also incorporate a diversity management strategy in the EA to avoid selecting highly similar parameter settings. Interested readers can find further information on this topic in the Supplementary File S2. 

\subsection*{Step 4: Generation of new parametrizations}~\\
Subsequently, the previously selected parametrizations are used to generate a set of new parameterizations that are evaluated in the next iteration. This process involves applying various techniques to create a new set of parameterizations, drawing inspiration from the process of meiosis. In the following section, we will discuss these criteria and how they contribute to the overall success of the evolutionary process. The number of generated settings will again vary based on which iteration the algorithm is in (see Table \ref{tab:parameter_values2}). The values given below are for iterations 11 onwards with a total of 300 settings generated.

\subsubsection*{Step 4.1: Selected parameter settings}~\\
Initially, all 30 previously chosen parameterizations are considered again in the next iteration. This criterion facilitates assessing the same parameter settings using a new random seed in each iteration, resulting in a more robust evaluation of their performance.

\subsubsection*{Step 4.2: Combination of selected parameter settings}~\\
Furthermore, 180 parameterizations are generated by combining two randomly chosen parental parameterizations in Step 3, denoted as \(X = (x_1, x_2, \ldots)\) and \(Y = (y_1, y_2, \ldots)\). By using these parents, a new parametrization \(Z = (z_1, z_2, \ldots)\) is created in each case.
\par
For continuous variables this is done by the use of a weighted average:

\begin{linenomath}
\begin{equation*}
\mathrm{z}_{i}\ = w{x}_{i} + (1-w){y}_{i} \quad \text{with} \quad w \sim U(0,1)
\end{equation*}
\end{linenomath}

For class variables, we randomly sample $z_i$ with an equal probability of belonging to either the class of $x_i$ or $y_i$. In our toy example, it is necessary to furthermore round and scale continuous variables to obtain integer numbers while fulfilling the budget constraint (see Step 1). 
\par

Following the combination process, we introduce small changes to the combined parametrizations, inspired by the process of allelic mutation in meiosis. For continuous parameters, the size of the mutation $t_i$ in parameter $i$ is sampled from a uniform distribution with the range determined by the variance of the parameter and a mutation occurring in each respective parameter with a probability of $p_{activ,i}$:
 
\begin{linenomath}
\begin{gather*} 
 Z_{mut} = (z_1 + m_{activ,1} t_1, z_2 + m_{activ,2} t_2, \ldots)\\
 \quad \text{with} \quad t_i \sim U(-2\sigma_{x_i}, 2\sigma_{x_i})\\
 \quad \text{and} \quad m_{activ,i} \sim B(0, p_{activ,i})\\
\end{gather*}
\end{linenomath}

here $\sigma_{x_i}$ denotes the standard deviation and is derived based on the empirical variance of the parameterizations of the current iterations. In our algorithm, in the first iteration, $p_{activ,i}$ is set to 0.2, indicating a 20\% chance of mutation for all parameters.
\par
In later iterations of the algorithm, it can make sense to avoid combinations of parameterization with different values for the class variables as these settings might not be compatible with each other anymore as non-class variables are adapted to work particularly well with the class variables. In our case, the probability of having two parameterizations with different class settings gets reduced as the algorithm progresses. By iteration 2, this probability is decreased by 20\%, and it keeps decreasing by 10\% per iteration until an 80\% reduction is reached. Similarly, the mutation rate $p_{activ,i}$ in class variables is reduced by 10\% in iteration 2 and reduced by 5\% per iteration until a reduction of 40\% is reached.

\subsubsection*{Step 4.3: Minor modifications of selected parameter setting}~\\
Lastly, we are considering the selected parent parameterization and applying minor changes / mutations to them (see Step 4.2). To avoid generating a setting already generated in Step 4.1, mutation rates $p_{activ,i}$ are increased to 0.3 and the sampling procedure is repeated in case no mutations are performed. 

\subsection*{Step 5: Convergence/ Optima/ Termination criteria)}~\\
To derive the optima, we suggest to first employing a kernel density estimation to determine which areas of the search space include sufficient coverage. We here propose to only consider those settings from the last five iterations with a value for the kernel density estimation above the 20\% quantile of these parametrizations, to avoid using parameterizations in sparsely sampled areas. For the estimation of the kernel density estimation, only simulations from the last five iterations are used.

\begin{linenomath}
\begin{dmath*}
f(y) = \frac{1}{n^p \cdot h_1 \cdots h_p} {\sum_i K(\frac{y_1-x_i,_1}{h_1}, \cdots,\frac{y_n-x_i,_p}{h_p})}
\end{dmath*}
\end{linenomath}

with $p$ being the number of parameters and using the empirical standard deviation in the last five iterations as the bandwidth $h_j$ and the use of a multivariate Gaussian kernel for $K$:

\begin{linenomath}
\begin{equation*}
K(y_1, \cdots, y_p) = K_1(y_1) \cdots K_p(x_p)
\end{equation*}
\end{linenomath}
with
\begin{linenomath}
\begin{equation*}
K_i(x) = \frac{1}{\sqrt{2\pi}}\exp\left(-\frac{x^2}{2}\right)
\end{equation*}
\end{linenomath}

Subsequently, the expected performance of all remaining parametrizations is estimated using a kernel regression (see Step 3.2 in \cite{hassanpour2023optimization}) using all simulations. The parametrization with the highest value based on the kernel regression is used as the optima.

\par
In our example, we performed 40 iterations without any termination criteria accessed. To define a general termination criteria, we propose to assess the optima from all previous iterations based on a kernel regression based on all simulations and if improvements for ten iterations are below a certain threshold terminate the pipeline. This threshold needs to be chosen depending on the specific optimization problem and is highly dependent on the desired precision of results \citep{jain2001termination, ghoreishi2017termination}. Given the time-consuming nature of simulating real breeding programs, we suggest users invest sufficient time in manually monitoring the algorithm's performance, e.g. by visual inspection. If users observe no substantial changes or improvements over successive iterations in both parameter settings and the objective function's value but also the computational cost arising from the algorithm, they may consider stopping the optimization process earlier. A possible alternative approach could be, to begin with a more conservative iteration limit and only increase it if the need arises after further evaluation.

\subsection*{Step 6: Final assessment of the optima}~\\
After the optimum is identified, the finally obtained breeding scheme is analyzed in-depth, as the kernel regression will naturally be biased in an optimum. For this, the suggested optimum is simulated a high number of times, e.g. in our case 100 replicates.

\subsection*{Modified parameter settings}~\\
Depending on the optimization problem and the iteration of the algorithm, adapting parameter settings can improve the convergence of the pipeline. This section is less about providing concrete values that work best universally but more about offering intuition on when and whether to deviate from the presented default and in which direction.
\par
Regarding the initial population size (Step 1), the goal should be to obtain a good coverage of the initial search space. Therefore, with more parameters or larger search intervals, it can make sense to increase the size of the initial set of parameterizations to a couple of thousand. In case simulations require a high computational load \citep{osti_790398, Piszcz.07082006}, it might be necessary to reduce the initial set of parameterizations. However, one should be aware that this will increase the risk of running into a local maximum \citep{BAJER2016294}.  
\par
To assess parameterizations more effectively, it may be advantageous to evaluate scenarios with multiple replicates within a single iteration (Step 2). This approach is particularly relevant for extremely small breeding programs and short time horizons where stochasticity can be a major factor in the evaluation. However, even for our small toy example, such extensive replication was not necessary. Replication can also be employed as a strategy to estimate the variability of the outcome of the simulation and then be integrated into the objective function to provide a more robust and accurate optimization solution.
\par
For selecting the best parameter settings (Step 3), we recommend selecting more parameterizations in the first couple of iterations, as there is still more diversity present and to avoid the loss of potentially promising settings. In our example, we selected 100 parameterizations in the first two iterations, 50 parameterizations in iterations 3-9, and 30 parameterizations afterward with similar splits between Steps 3.1, 3.2, and 3.3 (Table \ref{tab:parameter_values2}).
\par
For generating new parameter settings (Step 4), the same number of new parameterizations (300) in each iteration is generated. However, as Step 4.3 is mostly intended for fine-tuning already promising settings, this is not applied in the first few iterations, and more focus is given to Steps 4.1 and 4.2.
\par
As a refinement to Step 4, mutation rates can be adjusted based on the observed changes in each parameter during previous iterations. For example, a binary parameter that consistently shows superior results with one of the parameterizations should undergo less frequent mutation. For details on the approach used in our toy example, the interested reader is referred to Supplementary File S3.

\subsection*{Alternative Scenarios}
\subsubsection*{Scenario 2 - Reduced initial search space}~\\
In our previous study \citep{hassanpour2023optimization}, we identified optimal parameter settings (2368, 175, 19) for the three parameters considered. To assess the effectiveness of our EA algorithm, Scenario 2 tackles the same resource allocation problem as Scenario 1. However, we consider an initial search space that does not contain the optima to determine the ability of the EA to still identify the optima. For this we are considering the following two additional constraints in the initialization (Step 1):

\begin{linenomath}
\begin{gather*} 
\begin{aligned}
 300 &\leq x_2 \leq 500 \\
 15 &\leq x_3 \leq 25
\end{aligned}
\end{gather*}
\end{linenomath}

\subsubsection*{Scenario 3}~\\
For showcasing the optimization of a class variable, we extend Scenario 1 by introducing a binary variable, $x_4$. This variable represents a breeding strategy that leads to improved phenotyping causing a reduced residual variance and hence a higher heritability ($h^2$ = 0.32). For this study, we leave it open as to what this new breeding strategy is, but one could envision more uniform housing conditions, the use of electronic devices to measure physiological status, or large-scale collection of additional data like mid-infrared spectroscopy \citep{BOICHARD2012544, de2015housing}. Step 2 is accordingly adapted by sampling initial values for $x_4$ from a Bernoulli distribution \(x_4 \sim \mathrm{B}(0.5)\). We are here considering two versions of the scenario (3a / 3b) with varying additional costs of 1,000€ / 10€, respectively. Resulting in the new constraints:

\begin{linenomath}
\begin{gather*}
3a:  x_1(4000 + 1000x_4) + 3000x_2 - 10000000 \leq 0
\end{gather*}
\end{linenomath}
\begin{linenomath}
\begin{gather*}
3b:  x_1(4000 + 10x_4) + 3000x_2 - 10000000 \leq 0
\end{gather*}
\end{linenomath}

\subsection*{Snakemake}
The EA algorithm is iterative, involving multiple interdependent steps, where the computational demands for executing some steps are notably high, particularly the resource-intensive simulation of breeding programs in Step 2. To effectively address this computational challenge, the implementation of parallel processing in an efficient manner is crucial. 
\par
\begin{figure}[H] 
\centering
\includegraphics[width=3.8in, height=8.9in]{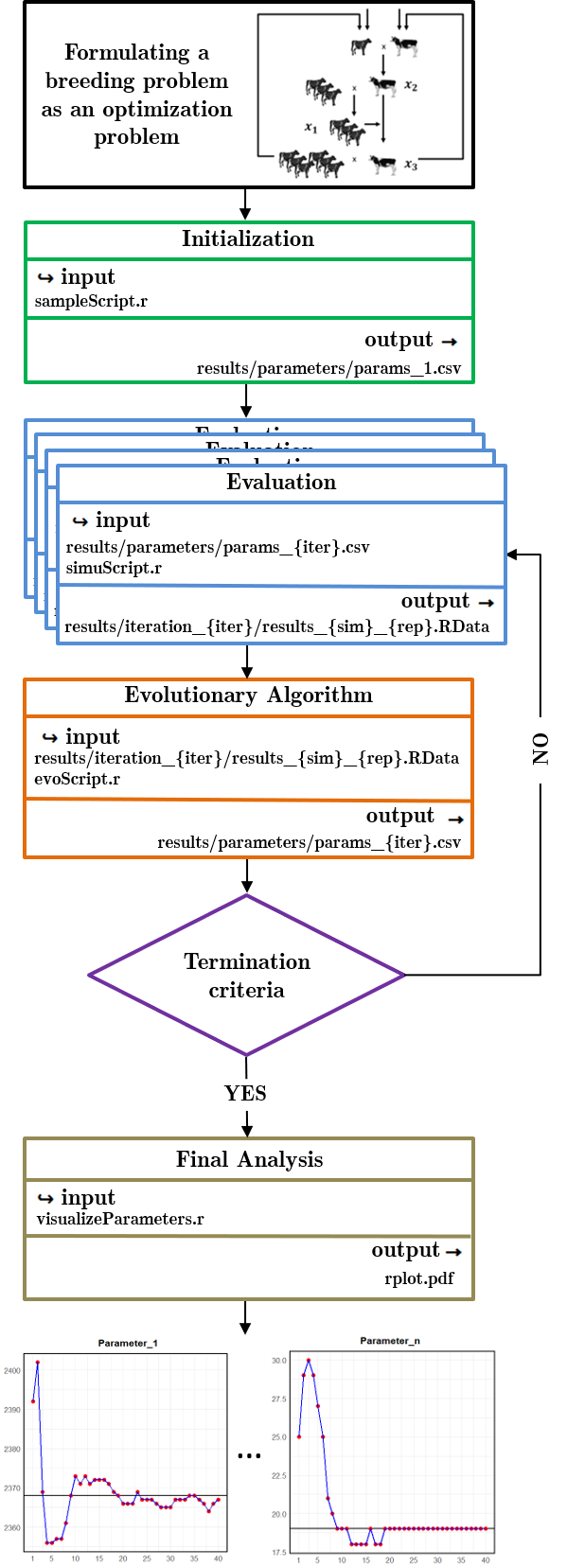}
\caption{Example visualization of the Snakemake workflow. Shown are the rule names defined and their input-output relationships.}%
\label{fig:figure2}
\end{figure}

For this purpose, our optimization pipeline makes use of the automation provided by the Snakemake workflow management system \citep{Molder.2021}. An illustrative representation of the Snakemake workflow for our evolutionary optimization model is provided in Figure \ref{fig:figure2}. Our Snakemake process uses four rules, corresponding to the steps of the algorithm, which are initialization (step 1), evaluation (step 2), evolutionary algorithm (steps 3, 4, and 5), and the final in-depth analysis of the obtain optima (step 6). 
\newpage
  
\par
To clarify, the individual simulations within step 2 are completely independent of each other and can easily be run in parallel using the built-in capabilities of Snakemake to seamlessly interact with various job scheduling systems (e.g., SLURM \url{https://slurm.schedmd.com/documentation.html}) on distributed hardware stacks, thus ensuring portability to a wide range of hardware setups. Those interested in configuring this setup can refer to the Snakemake plugin catalog at \url{https://snakemake.github.io/snakemake-plugin-catalog/index.html}. This catalog provides a starting point for configuring Snakemake to work with various cluster schedulers, ensuring optimal distribution and execution of multiple tasks across the computing cluster.

\subsection*{Computer hardware}
All tests were executed on a server cluster with Intel Platinum 9242 (2X48 core 2.3 GHz) CPUs using Snakemake toolkit version 7.21.0, which was configured to distribute single jobs via a SLURM scheduler to the backends of the cluster. Simulations were conducted on single nodes using a single core per simulation, taking approximately 15 minutes, and peak memory usage of 5 GB RAM per simulation. The computing time of all other steps combined increases approximately linearly in the number of iterations, but even in iteration 40 only took a negligible seven seconds. 

\section*{Results} 
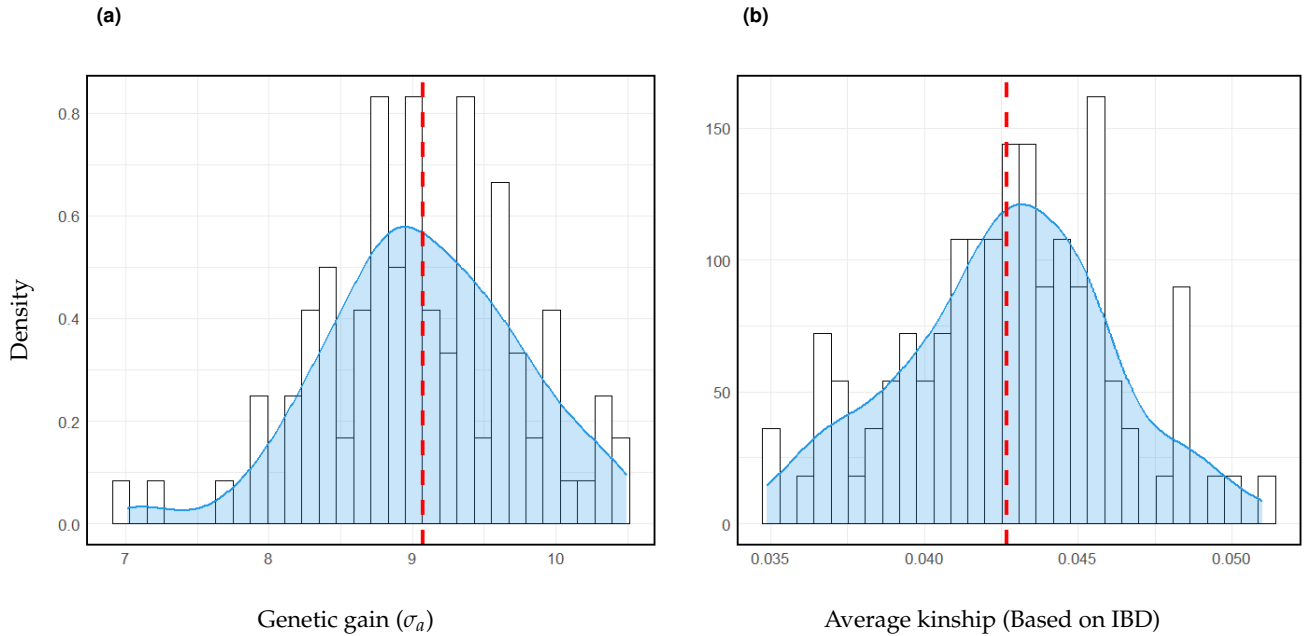
\begin{figure*}
    \centering
    \begin{tikzpicture}
        \matrix (fig) [matrix of nodes]{
            |[text width=2.6in]| {\subcaption{}\label{fig:figureS_1a}}
            &
            |[text width=2.6in]| {\subcaption{}\label{fig:figureS_1b}}
            \\
            \includegraphics[width=3.3in]{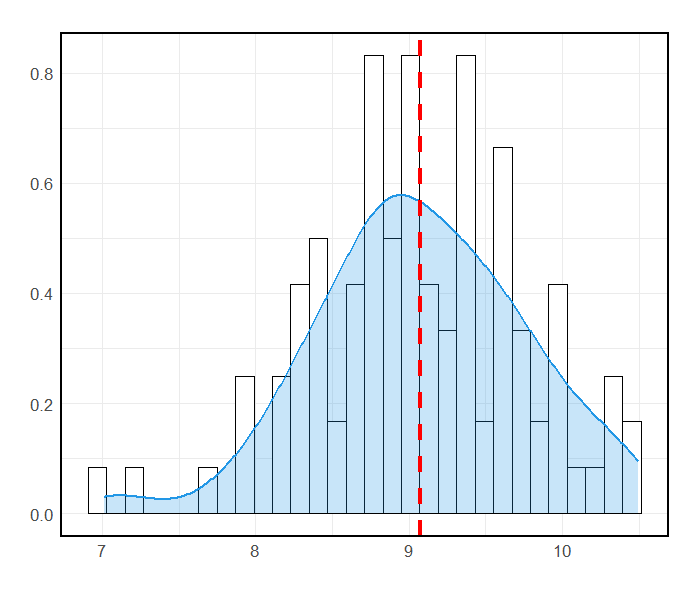}  
            &
            \includegraphics[width=3.3in]{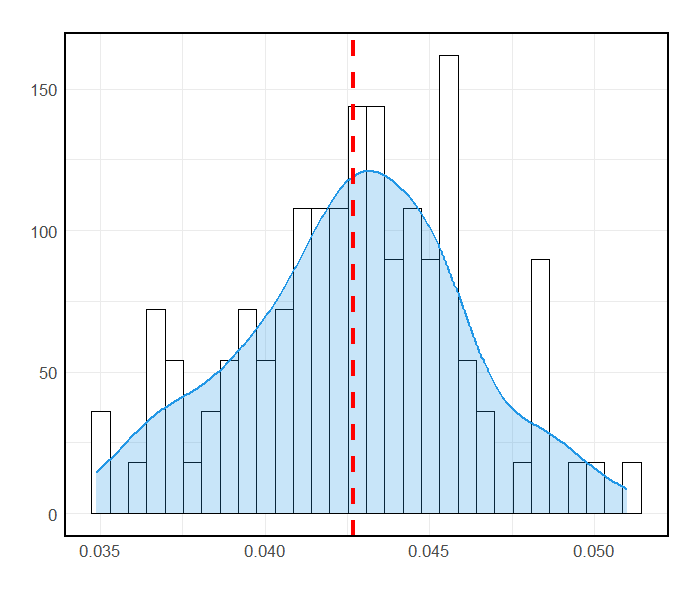}
            \\                 
        };
        \node[below=7.5cm of fig-1-1] {Genetic gain ($\sigma_a$)};
        \node[below=7.5cm of fig-1-2] {Average kinship (Based on IBD)};
        \path (fig-2-1.south west)  -- (fig-2-1.north west) node[midway,midway,sloped]{Density};
         \path (fig-2-1.south)  -- (fig-2-1.south) node[midway,below, yshift=0.5cm]{};
    \end{tikzpicture}
    \captionsetup{width=0.9\textwidth}
    \caption{Realization of expected outcome of the formulated objective function for Scenario 1 based on 100 replicates in (\ref{fig:figureS_1a}) for genetic gain ($\sigma_a$), (\ref{fig:figureS_1b}) for average kinship (Based on IBD). The red dashed line represents the mean value, while the blue shaded area shows the probability density.}
\end{figure*}

Application of our evolutionary pipeline to the optimization problem formulated in Scenario 1 suggests a final optimum of 2368 test daughters and 175 test bulls, of which 19 test bulls are selected ($x_1$ = 2368, $x_2$ = 175, $x_3$ = 19) with an expected outcome for the target function of 107.041, with a genetic gain of 9.07 genetic standard deviations (Figure \ref{fig:figureS_1a}) and an increase of the inbreeding level of 0.0426 (Figure \ref{fig:figureS_1b}) after 10 generations based on the averages of 100 replicates of this scenario. 

\par
\begin{figure*}
        \centering
        \begin{tikzpicture}
            \matrix (fig) [matrix of nodes]{
                |[text width=2.6in]| {\subcaption{Number of Test Daughters ($x_1$)}\label{fig:figure3_1}}
                &
                |[text width=2.6in]| {\subcaption{Number of Test Bulls ($x_2$)}\label{fig:figure3_2}}
                \\
                \includegraphics[width=3.3in]{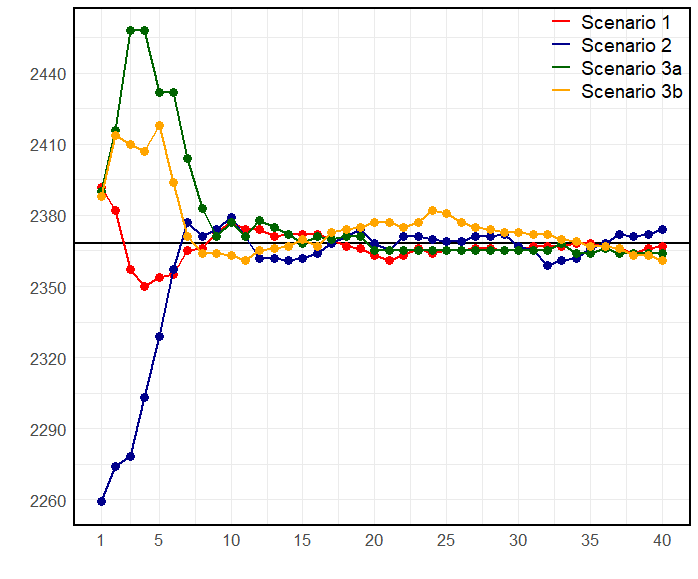}  
                &
                \includegraphics[width=3.3in]{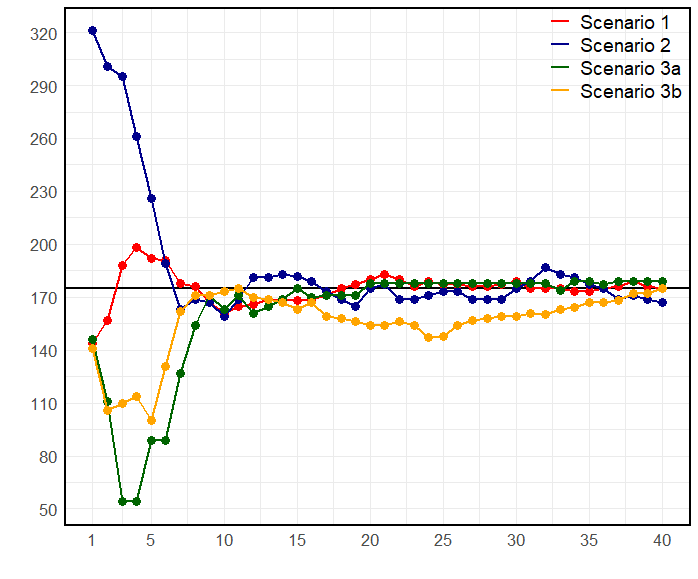}
                \\ 
                 |[text width=2.6in]| {\subcaption{Number of Selected Sires ($x_3$)}\label{fig:figure3_3}}
                &
                |[text width=2.6in]| {\subcaption{Binary Variable ($x_4$)}\label{fig:figure3_4}}
                \\
                \includegraphics[width=3.3in]{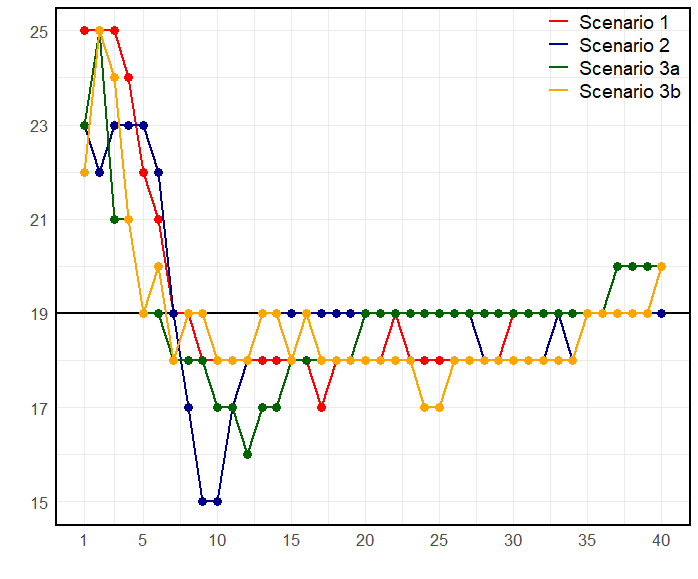} 
                &
                \includegraphics[width=3.3in]{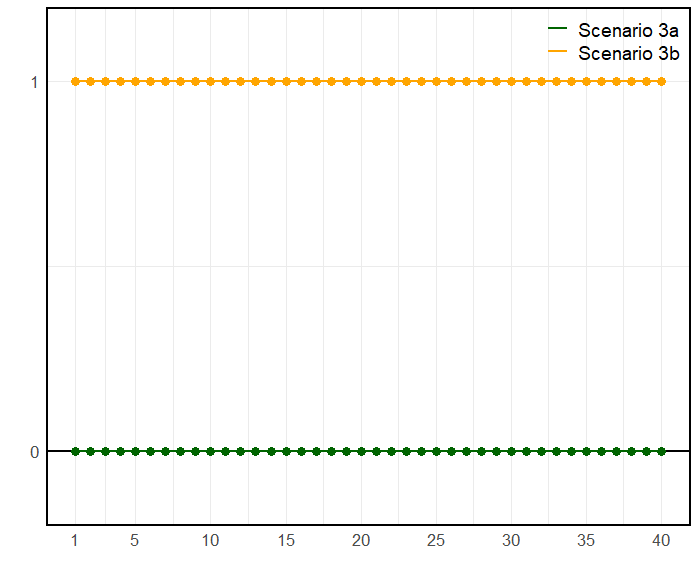} 
                \\
            };

            \path (fig-4-1.south west)  -- (fig-2-1.north west) node[above,midway,sloped,yshift=0.5cm]{Optimal Values};
            \path (fig-4-2.south)  -- (fig-4-1.south) node[midway,below]{Number of Iterations};
        \end{tikzpicture}
            \captionsetup{width=0.9\textwidth}
        \caption{Suggested optima for the individual parameters of the breeding program design for the number of test daughters (\ref{fig:figure3_1}), test bulls (\ref{fig:figure3_2}), and selected sires (\ref{fig:figure3_3}), as well as the binary variable (\ref{fig:figure3_4}). The black horizontal line represents the estimated optima through comprehensive exploration, achieved by conducting over 100,000 simulations utilizing kernel regression \citep{hassanpour2023optimization}}
        
    \end{figure*}
\begin{figure*}
    \centering
    \begin{tikzpicture}    
        \matrix (fig) [matrix of nodes]{
          |[text width=1.85in]| {}
            \\
            \includegraphics[width=5in, height=3in]{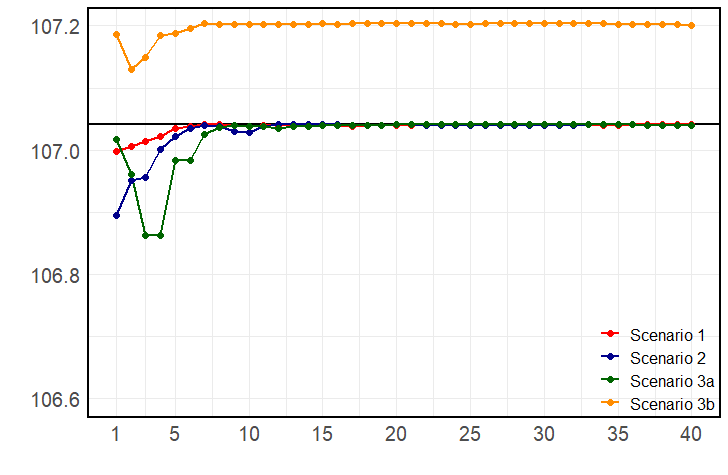}  
            \\
        };
        \path (fig-2-1.south west)  -- (fig-2-1.north west) node[midway,midway,sloped]{Optimal Value};
         \path (fig-2-1.south)  -- (fig-2-1.south) node[midway,below, xshift=1cm]{Number of Iterations};
    \end{tikzpicture}
        \captionsetup{}
    \caption{Performance of the suggested optima after each iteration assessed using kernel regression based on all simulations in Scenario 1. The black horizontal line shows the estimated optima obtained from \citet{hassanpour2023optimization}.\label{fig:figure4}} 
\end{figure*}

All three individual parameters very quickly reached values close to the finally suggested optima (Figure \ref{fig:figure3_1}:\ref{fig:figure3_3}). When evaluating the suggested optima per iteration based on all simulations conducted (to avoid effects of reduced bandwidth over iterations) even after seven iterations a value of 107.038 for the target function based on kernel regression was obtained (Figure \ref{fig:figure4}), despite kernel regression by design being downward biased in the optimum. 

\begin{figure*}[h!]
    \centering
    \pgfmathsetlength{\imagewidth}{\linewidth}%
    \pgfmathsetlength{\imagescale}{\imagewidth/524}%
    
    \begin{tikzpicture}[x=\imagescale, y=-\imagescale]
        \node[anchor=north east] (img1) at (0,0) {\includegraphics[width=0.8\imagewidth]{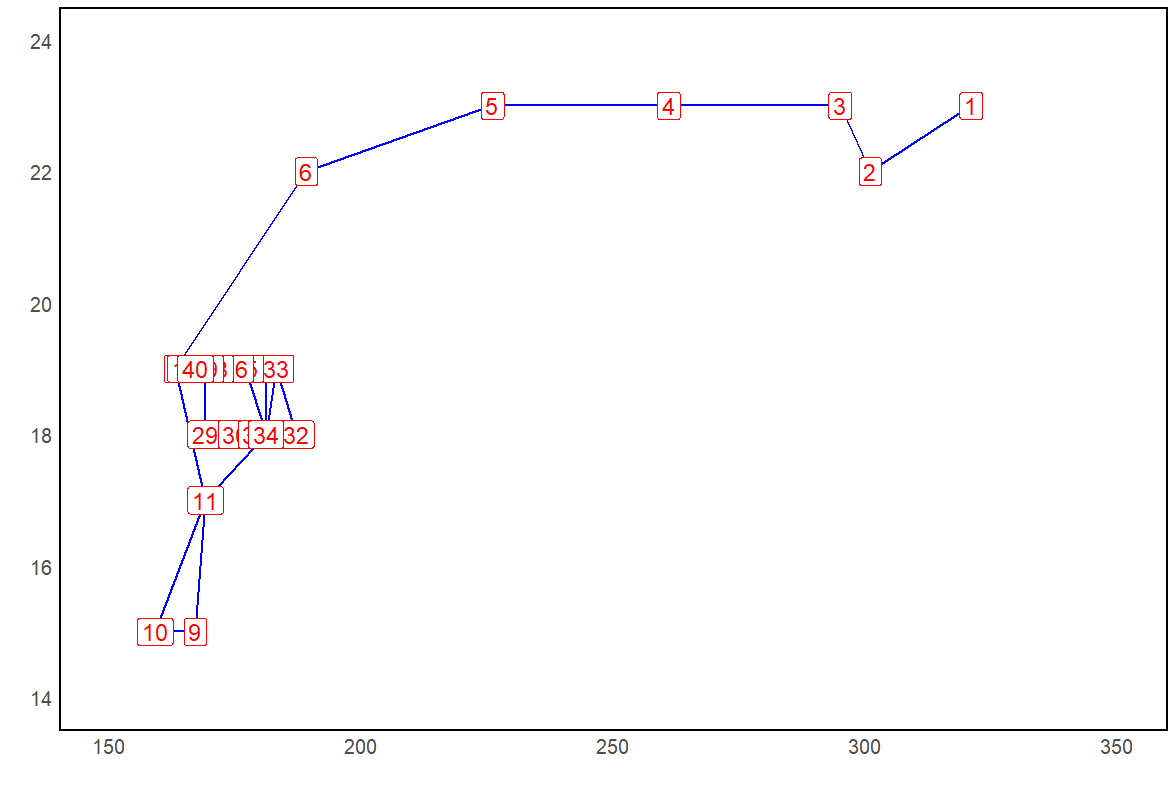}};
        \node[anchor=north east] (img2) at (-10,116) {\includegraphics[width=0.45\imagewidth]{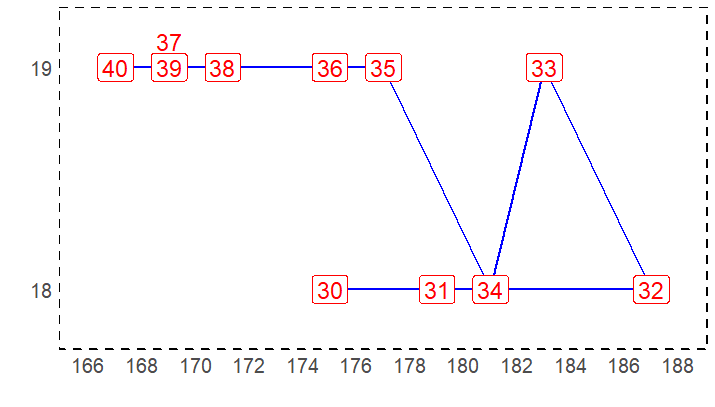}};
        
         \coordinate (dashed-start) at (-310,162);
         \coordinate (dashed-end) at (-230,233);
         
         \coordinate (dashed2-start) at (-315,130);
         \coordinate (dashed2-end) at (-230,122);

        \draw[] (img1) node[above,xshift=0.5cm, yshift=-5.5cm] {Number of Test Bulls ($x_2$)};
        \draw[] (img1) node[left,xshift=-7.5cm, yshift=2.2cm, rotate=90] {Number of Selected Sires ($x_3$)};
        \draw [black, dashed] (dashed-start) -- (dashed-end);
        \draw [black, dashed] (dashed2-start) -- (dashed2-end);
        \draw (-123,142) circle (7pt);
    \end{tikzpicture}
    
    \captionsetup{width=0.9\textwidth}
    \caption{Suggested optima across iterations for Scenario 2. Red labels denote iteration numbers, while the blue line illustrates the iterative pathway. The dashed segment zooms in on the overlapping iterations within the optimal range from iteration 30 to 40. The black circle denotes the reference point for optimal parameter settings obtained through kernel regression, involving over 100,000 simulations.}
    \label{fig:figure5}
\end{figure*}

In Scenario 2, basically the same optimum is obtained after 40 iterations, with $x_1$ = 2374, $x_2$ = 167, and $x_3$ = 19. In the initial stages of iteration, the optimal solutions provided in the first few iterations perform slightly below the optimum of Scenario 1. However, after seven iterations, the values of the target function for the suggested optima are now comparable to those of Scenario 1. (Figure \ref{fig:figure4}). The individual parameters exhibit more substantial changes during the initial iterations, indicating that the use of more test bulls could be beneficial, due to limitations in the initial search space. Similarly, a higher number of sires is selected ($x_3$) to use a similar selection intensity. In the first ten iterations of Scenario 2, the primary emphasis is on quickly bringing $x_2$ into its optimal range due to the higher overall impact of $x_2$ on the optimum of the formulated objective function. Overall, more change in the individual parameters is observed with $x_3$ in iteration 10 being as low as 15 (Figure \ref{fig:figure3_3}). A detailed overview of the changes in $x_2$ and $x_3$ is given in Figure \ref{fig:figure5}.

\begin{figure*}[htbp]
        \centering
        \begin{tikzpicture}
            \matrix (fig) [matrix of nodes]{
                |[text width=1.8in]| {\subcaption{Initial population}\label{fig:figure6_1}}
                &
                |[text width=1.8in]| {\subcaption{Iteration 5}\label{fig:figure6_2}}
                &
                |[text width=1.8in]| {\subcaption{Iteration 10}\label{fig:figure6_3}}
                \\
                \includegraphics[width=2.2in]{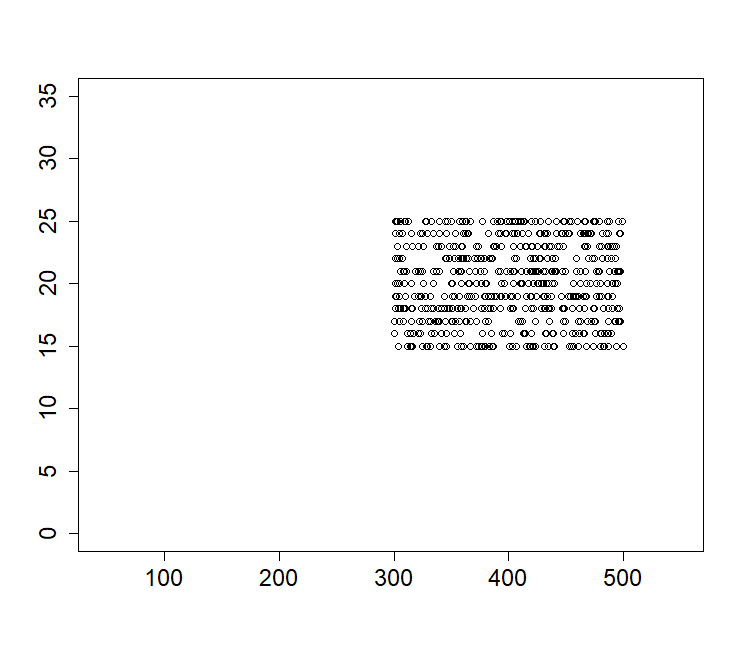}  
                &
                \includegraphics[width=2.2in]{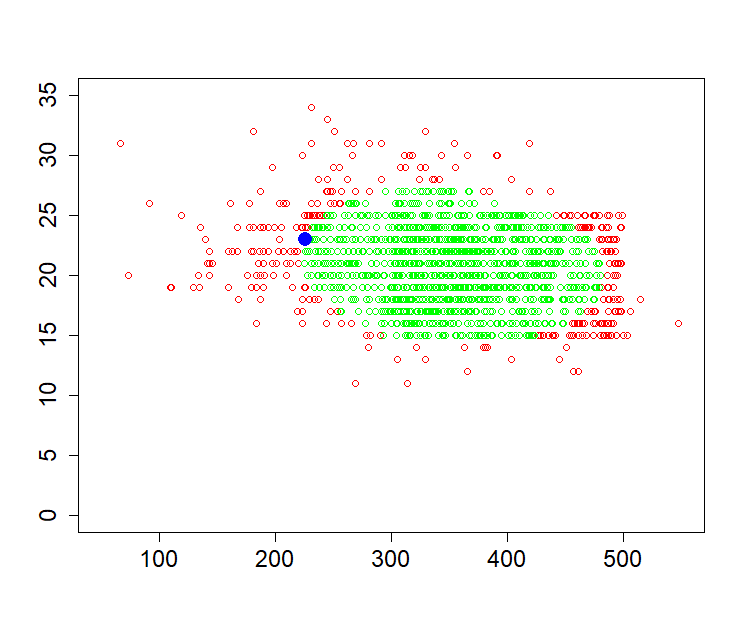}
                &
                \includegraphics[width=2.2in]{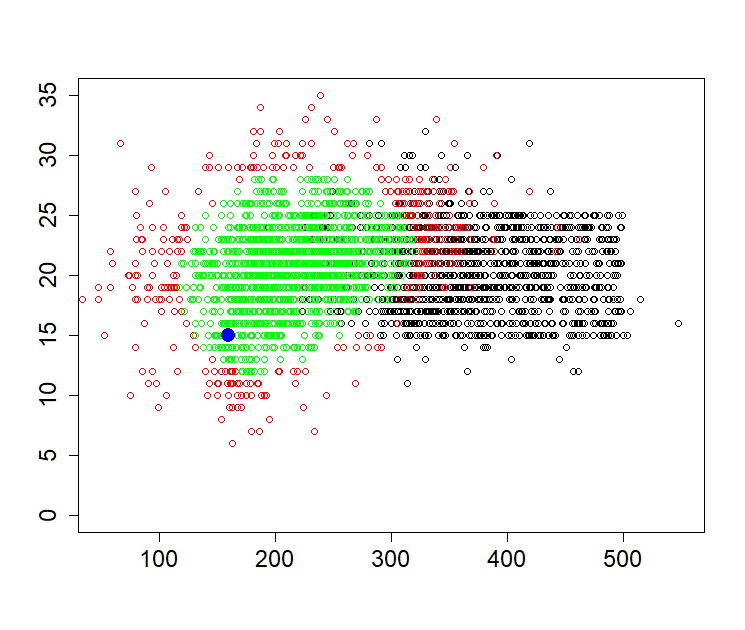}
                \\ 
                |[text width=1.8in]| {\subcaption{Iteration 15}\label{fig:figure6_4}}
                &
                |[text width=1.8in]| {\subcaption{Iteration 20}\label{fig:figure6_5}}
                &
                |[text width=1.8in]| {\subcaption{Iteration 25}\label{fig:figure6_6}}
                \\
                \includegraphics[width=2.2in]{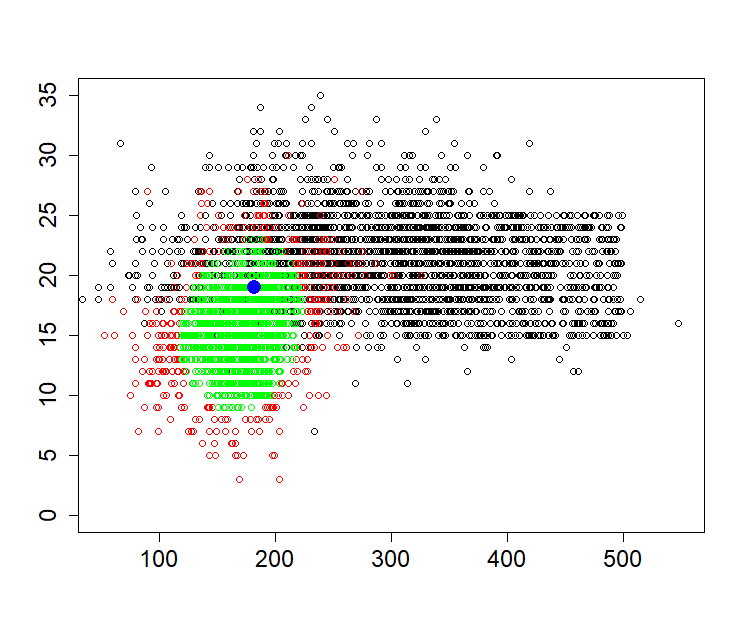} 
                &
                \includegraphics[width=2.2in]{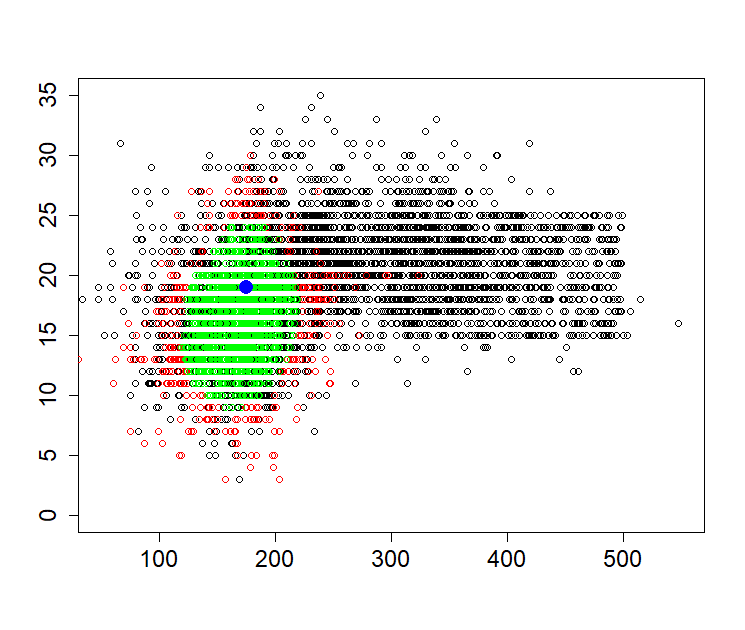} 
                &
                \includegraphics[width=2.2in]{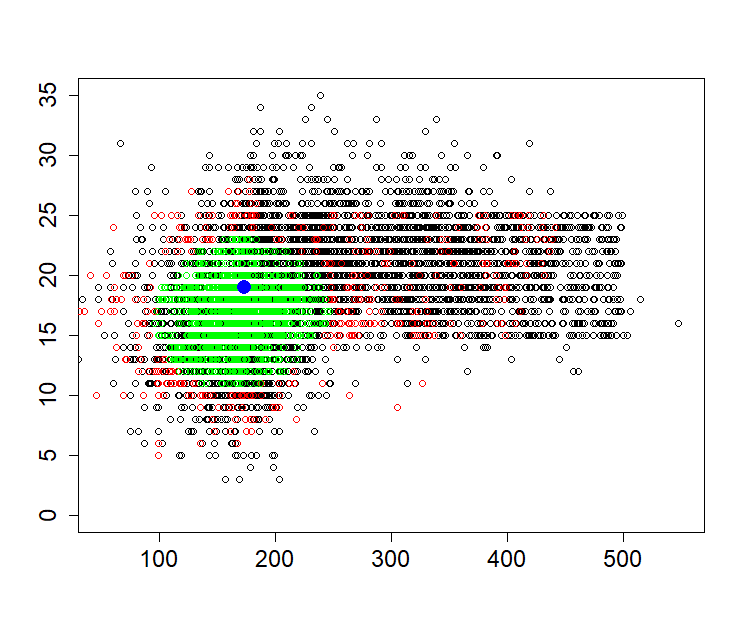} 
                \\
                |[text width=1.8in]| {\subcaption{Iteration 30}\label{fig:figure6_7}}
                &
                |[text width=1.8in]| {\subcaption{Iteration 35}\label{fig:figure6_8}}
                &
                |[text width=1.8in]| {\subcaption{Iteration 40}\label{fig:figure6_9}}
                \\
                \includegraphics[width=2.2in]{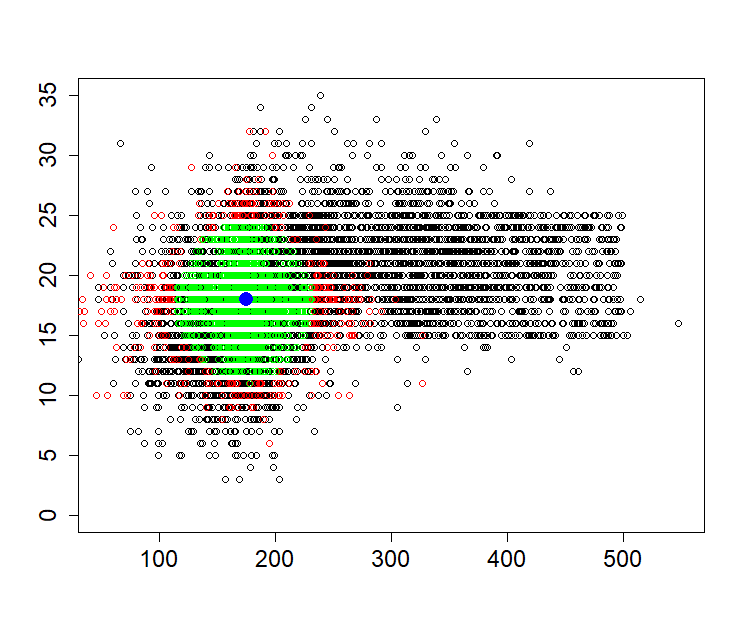} 
                &
                \includegraphics[width=2.2in]{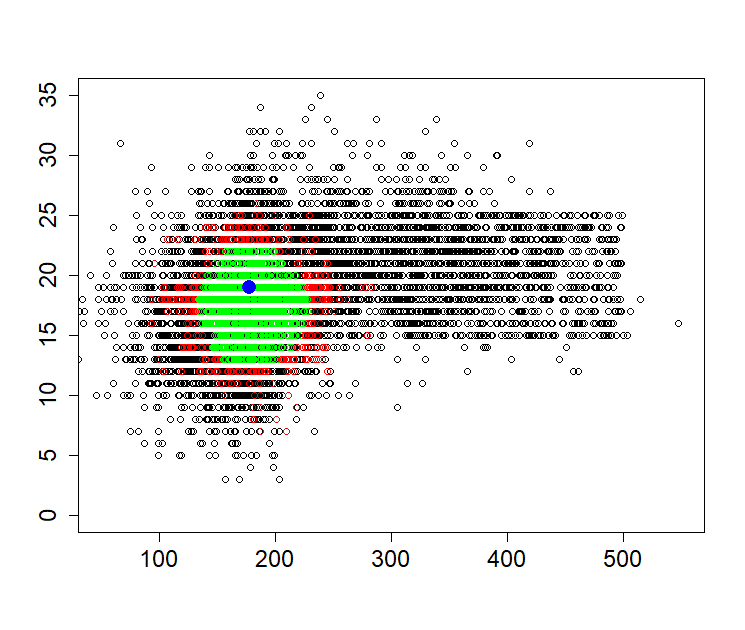}
                &
                \includegraphics[width=2.2in]{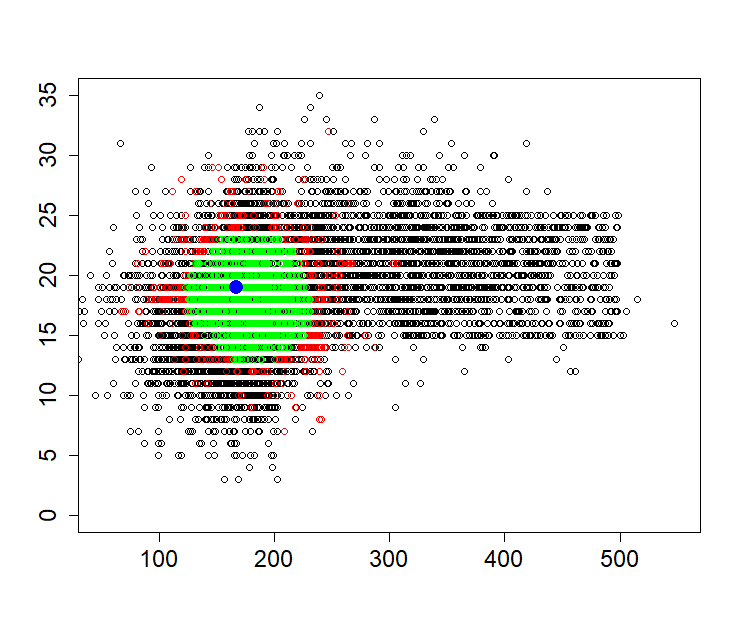}
                \\
            };

            \path (fig-4-1.south west)  -- (fig-4-1.north west) node[,midway,sloped,yshift=0.7cm]{Number of Selected Sires ($x_3$)};
            \path (fig-4-2.south)  -- (fig-4-1.south) node[midway,below,xshift=3cm,yshift=-6cm]{Number of Test Bulls ($x_2$)};
        \end{tikzpicture}
        \captionsetup{width=1\textwidth}
        \caption{Estimates of the optimum values in Scenario 2: (\ref{fig:figure6_1}) Initial population, (\ref{fig:figure6_2}) Iteration 5, (\ref{fig:figure6_3}) Iteration 10, (\ref{fig:figure6_4}) Iteration 15, (\ref{fig:figure6_5}) Iteration 20, (\ref{fig:figure6_6}) Iteration 25, (\ref{fig:figure6_7}) Iteration 30, (\ref{fig:figure6_8}) Iteration 35, and (\ref{fig:figure6_6}) Iteration 40. The black points in the illustration represent simulations from more than five iterations back that are no considered as potential optima. Red points represent simulations that were excluded as optima due a low kernel density estimation. Green points represent candidate optima from the kernel density estimation with the blue dot representing the finally chosen parameterization.}
    \end{figure*}

Although parametrizations with smaller $x_2$ values do exist in early iterations (Figure \ref{fig:figure6_2}), these are not considered in deriving the optima (Step 5) due to the low kernel density (red area). In later iterations the area of solutions considered as the optima more and more shifts towards the area with the expected optima (green area, Figure \ref{fig:figure6_3}:\ref{fig:figure6_9}).

In Scenario 3a basically the same optima as in Scenarios 1 \& 2 is obtained with the binary not being active ($x = (2365,179,20,0)$). In terms of convergence speed, the individual parameters exhibit more variation in the early iterations, particularly with iterations 4 and 5. This is because the stochasticity in the evaluation of the target function (Step 2) leads to the suggestion of optima that are later identified as poor solutions, as shown in Figures \ref{fig:figure3_1}:\ref{fig:figure3_3} and \ref{fig:figure4}.

\par
In contrast, the binary is active in Scenario 3b which allows for a higher overall value of the target function of 107.200 that represents a statistically significant improvement based on a t-test \citep{Student1908} ($p < 0.00663$). Due to the higher overall housing costs, the number of cows and bulls is slightly decreased with a finally suggested optimum of $x  = (2361,175,20,1)$.

\par
\begin{figure*}
        \centering
        \begin{tikzpicture}
            \matrix (fig) [matrix of nodes]{
                |[text width=2.6in]| {\subcaption{Scenario 3a}\label{fig:figureS_2a}}
                &
                |[text width=2.6in]| {\subcaption{Scenario 3b}\label{fig:figureS_2b}}
                \\
                \includegraphics[width=3.3in]{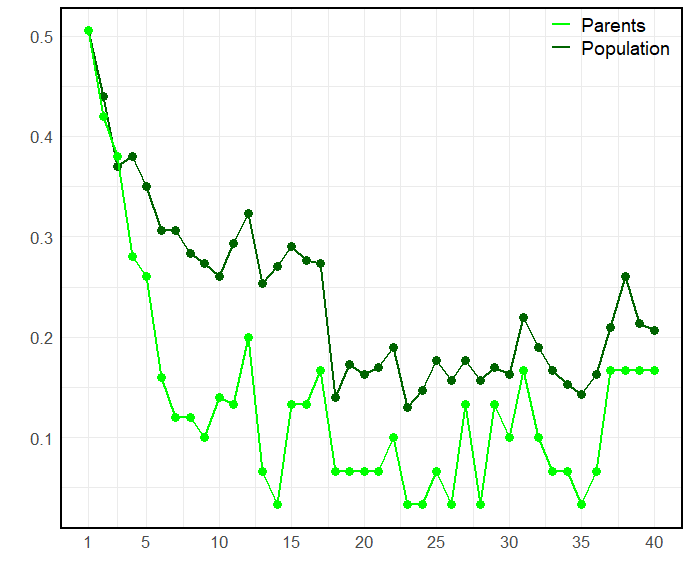}  
                &
                \includegraphics[width=3.3in]{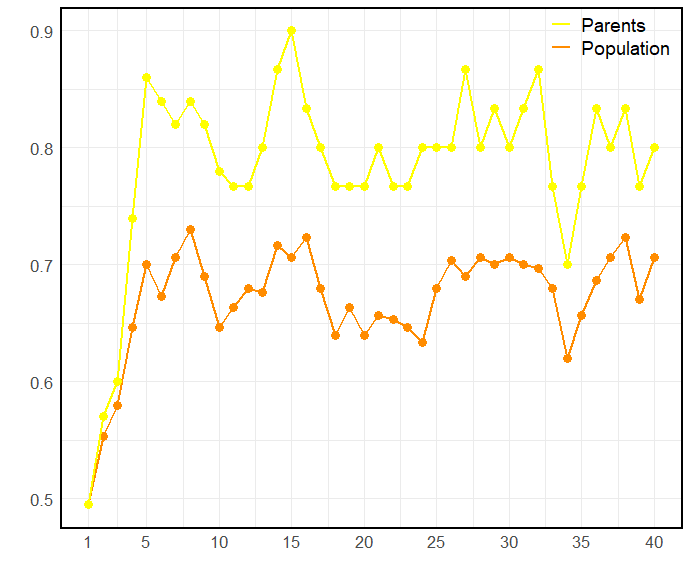}
                \\                 
            };

            \path (fig-4-1.south west)  -- (fig-2-1.north west) node[above,midway,sloped,yshift=0.1cm]{Share of Binary Variable};
            \path (fig-4-2.south)  -- (fig-4-1.south) node[midway,below, xshift=3cm,yshift=-1.5cm]{Number of Iterations};
        \end{tikzpicture}
            \captionsetup{width=0.9\textwidth}
        \caption{The share of a binary variable is shown in each iteration, to increase the heritability of phenotyping in (\ref{fig:figureS_2a}) for Scenario 3a with green being the share of parents and dark green being the share of population, (\ref{fig:figureS_2b}) for Scenario 3b with yellow being the share of parents and dark orange being the share of population.}
     
    \end{figure*}

Regarding the binary parameter, in Scenario 3a, the mutation rates are reduced to half from iteration 17 onwards. In contrast, the mutation rates remain high for the entire 40 iterations in Scenario 3b. In both scenarios, the selected parameterizations (Step 3) have a higher share of the favorable binary compared to the overall population. Over the entire simulation, 18\% of the parameterizations in Scenario 3a and 25\% in Scenario 3b have the alternative binary setting, as shown in Figures \ref{fig:figureS_2a} and \ref{fig:figureS_2b}, respectively.

\section*{Discussion} 
In this study, we introduce a new EA framework that can optimize breeding program design by handling both categorical and continuous variables. This framework enables the joint optimization of multiple design parameters in a computationally efficient way, making it suitable for breeding program design. 
\par
The here-developed pipeline represents a much-enhanced version of the kernel regression pipeline suggested in \citet{hassanpour2023optimization}. The iterative nature of the EA proves advantageous, as each iteration generates more data (simulations), which in turn enhances the reliability and efficiency of kernel regression in identifying suitable parameterizations for a breeding scheme. Furthermore, kernel regression remains a crucial component in addressing the challenges of optimization problems with stochastic noise in the target function evaluation, a common issue when utilizing stochastic simulations \citep{hart1996optimization,liang2000evolutionary}.

\par
The developed pipeline provides a lot of flexibility to easily adapt parts of the algorithm to improve its efficiency but also the breeding program design itself. Nonetheless, models also provide robustness, as shown in Scenario 1, 2, and 3a with independent runs obtaining very similar final results. Note here that the term convergence should in this context be used with caution, as the finally obtained optima can gradually change with decreasing bandwidth of kernel regression. Therefore the suggested "optima" will most likely not be the exact optima but at least be very close to it. Note that the stochasticity in the evaluation of the target function will naturally cause minor deviations between runs and although differences in solutions will exist, practically, suggested optima between the three scenarios are very similar. The suggested optima in Scenario 1 exactly matching the optima from \citet{hassanpour2023optimization} should therefore mostly be seen as a coincidence, but not as a sign of exact convergence. 
\par
As a recommendation, users should regularly track the level of improvement of the objective function during the optimization process. It is important to recognize that extensive iterations, such as running the optimization process for additional cycles, may incur more computational expense than the profit procured from a small gain per iteration. This gain may not be deemed substantial, considering the inherent stochasticity of the entire process and the likelihood of simulation bias.

\par

The robustness of the algorithm is particularly highlighted by Scenario 2, demonstrating that the EA algorithm can find optimal solutions even outside of the initial search space. This is an advantage over our previously established kernel regression method \citep{hassanpour2023optimization}, which relies on predefined parameter bounds and cannot dynamically adapt its search space. As such, it falls short in terms of automation and efficiency. Naturally, the choice of suited parameters and design space in the EA will make convergence both quicker and more reliable \citep{RAHNAMAYAN20071605,Kazimipour.2014}. Even Scenario 1 could have easily been improved in terms of required computations by not considering cases of more than 250 test bulls in the initialization. Nonetheless particularly in more complex optimization problems covering a wide range of the parameter space should usually be the priority \citep{zaharie2017revisiting}. Particularly with a higher number of parameters, this should reduce the risk of running into local maxima. To avoid running into local optima, one could for example extend the generation of new parametrizations (Step 4) by randomly sampling parametrizations similar to the initialization. Most of the simulations generated this way will be highly explorative with a low likelihood of providing good solutions and unnecessarily increase computational cost.
\par
Although the use of a fixed termination criterion and assessment of the state of the pipeline is often desirable, we strongly recommend also relying on manual human assessment, at least in support. Visual inspection of the change in target function and individual parameters is a common practice with evolutionary algorithms \citep{Almeida_2015}. In many cases, fine-tuning the parameters associated with termination criteria relies often on an iterative, trial-and-error approach \citep{jain2001termination}. By integrating Snakemake into our EA framework, we enhanced the flexibility in determining when to stop the algorithm without the need to rerun initial iterations. This is possible because a Snakemake process can be paused and then re-evaluated, deciding independently which steps need to be rerun, run additionally, or which results can be reused. In some instances, even if the value of the objective function remains relatively stable, there might be variation in individual parameter settings across iterations \citep{Moscato.1989}. This has practical implications in real-world scenarios, and it enables breeders to potentially allocate resources differently or achieve the same outcome through alternative scenarios. In this regard, the definition of a suitable target function is of major importance, as from practical experience defining such an objective function is not easy or at minimum very abstract. Here, visual manual human assessment can also help to check if the suggested optima from the EA is not only in the defined search space but also in the realm of solutions a breeder would grant reasonable / doable.

\par

For practical breeding, it would also be conceivable to run the EA pipeline multiple times, potentially with different target functions to then perform an in-depth analysis to calculate key characteristics (genetic gain, inbreeding, etc.)  from these optima and pick the preferred solution. By this, the abstract concept of a target function can be replaced with a practical choice between which combination of genetic gain / inbreeding is the most desirable. Different target functions could for example be generated by using different weightings of genetic diversity and short/long-term genetic gain.

\par

Assessing EAs's performance in terms of speed and computational effort is broadly defined to include various sensible metrics, such as the number of iterations, CPU time, or any similar indicators \citep{eiben2011evolutionary}, with the number of simulations required being the main driver of computational load in our pipeline. The efficacy of our EA framework in reducing computational resources is highlighted through its performance across all scenarios. For example, in Scenario 1 only 2400 simulations were required to achieve similar results as compared to our previous kernel regression approach which relied on more than 100,000 simulations \citep{hassanpour2023optimization}. This demonstrates a considerable reduction in computational effort, with the algorithm achieving a more than 40-fold decrease compared to the kernel regression method. This reduction holds even with a less-than-ideal initial search space or when adding a binary decision variable and therefore emphasizes the practical advantages of EA in optimizing large-scale breeding program designs. Additionally, scaling for a higher number of parameters should be greatly improved as the traditional kernel regression scales exponentially in the number of assessed parameters.

\par

In the context of economic considerations for optimization strategies, a crucial aspect involves evaluating the costs against potential benefits. For our toy example, with a cloud computing service charging around 0.012€ per CPU core hour and 0.012€ per 6 GB of memory/hour \url{https://hpc.ut.ee/pricing/calculate-costs}, the total per-job cost, combining core hours and RAM usage, is 0.55 cent per simulations. Running 40 iterations with 12.300 simulations would therefore result in a cost of 67.65€. In most commercial industrial practices, the focus of optimization lies in finding the best solution within a specific operating region or parameter space of interest that meets cost-effectiveness criteria and generates profits within a specified timeframe.
\par
In the process of optimizing breeding scheme design, \citet{jannink2023insight} showed that there are difficulties when using Bayesian optimization to allocate budgets effectively in breeding schemes. One of the limitations faced in this investigation and our previous work \citep{hassanpour2023optimization} is the lack of support for class variables. Our EA strategy helps addressing the computational intensity associated with continuous optimization problems involving class variables. These problems can be computationally intensive not only due to their combinatorial nature but also due to the increase in the number of possible outcomes \citep{pelamatti2018deal}. Particularly with a high number of class variables, the total number of combinations will rapidly increase to \(k^n\) for \(n\) class variables that all can take \(k\) values.

\par 
We would therefore strongly recommend using as low a number of class variables as possible. For example, in the considered Scenario 3b, the additional cost of improved phenotyping was extremely low, which from a human perspective makes it quite obvious to spend this additional money, but an optimization algorithm would not recognize this. As the resulting improvement for the target function is low, the overall upside is low, and high overall stochasticity in the evaluation is observed, the EA for all 40 iterations considered both binary settings. On the contrary, more substantial differences, such as an increase in the costs of phenotyping of 1000 Euro for a marginal improvement in precision, are easily detectable by the algorithm to be unsuitable. Therefore, we recommend that when a design decision is clear from theory or intuition, consider fixing the class variable or running separate optimization pipelines for both binary settings to simplify the process.

\par 

Furthermore, \citet{jannink2023insight} reported high variability in the outcomes of different runs of the Bayesian optimization pipeline depending on small changes such as input genotypes and therefore lacking the ability to draw general conclusions from the obtained results. In our case, input genotypes and trait architecture were randomly sampled for each simulation. As very similar optima were obtained in the scenarios that should have the same optima (Scenarios 1, 2, 3b) this should be a strong indicator of the generality and stability of the approach.

\section*{Conclusion}
In conclusion, our study presents a new optimization framework using an EA that integrates a local search approach based on a kernel regression model. Our framework shows superior optimization efficiency to existing approaches and is applicable to both classes and continuous variables, hereby, enabling breeders to explore a wider range of scenarios compared to traditional methods \citep{Bancic2023}. The results across all problems indicate that our proposed framework produces robust optima at reduced computation cost. This study demonstrates that the EA algorithm consistently converges towards a common optimal solution, showcasing its robustness and ability to identify globally optimal or near-optimal configurations. The algorithm's superior convergence speed, solution diversity, balance between exploitation and exploration, and robustness to stochasticity highlight its potential for larger breeding optimization tasks. The adaptable nature of our proposed framework makes it not only suitable for various future projects but also ensures flexibility in accommodating different breeding program designs. Users can easily modify, extend, or replace steps and adjust parameter choices as necessary. Thus, our framework supports optimization strategies that adjust to changing needs in breeding programs.

\section*{Data availability}
The presented evolutionary pipeline is patent pending under application number EP24164947.4 and EP24188636.5. Patent applicants are BASF Agricultural Solutions Seed US LLC and Georg-August-Universitaet Goettingen. Inventors are Torsten Pook, Azadeh Hassanpour, Johannes Geibel, and Antje Rohde. Academic, non-commercial use is possible under a public license with details given at \url{https://github.com/AHassanpour88/Evolutionary_Snakemake/blob/main/License.md}.

\section*{Acknowledgments}
We acknowledge the computational support from the Scientific Compute Cluster at GWDG, the joint data center of the Max Planck Society for the Advancement of Science (MPG), and the University of Goettingen.

\section*{Funding}
This research was supported by the BASF Belgium Coordination Center CommV.

\section*{Conflicts of interest}
The authors declare that they have competing interests related to this work. TP, AH, AR \& JG are inventors of the associated patent applications EP24164947.4 and EP24188636.5. The competing interest declared here does not alter the authors' adherence to all Genetics policies on sharing data and materials.

\bibliography{library}

\clearpage
\onecolumn
\section*{Supplementary Materials}\label{sec:num}
\renewcommand{\thefigure}{S~\arabic{figure}}
\renewcommand{\thetable}{S~\arabic{table}}
\setcounter{figure}{0}
\setcounter{table}{0}

\subsection*{File S1}
All scripts of the evolutionary pipeline are available at \url{https://github.com/AHassanpour88/Evolutionary_Snakemake/blob/main}.

\subsection*{File S2}
\subsubsection*{Dissimilarity criterion}~\\
The iterative selection process assesses the Euclidean distance ($d$) between a candidate and the previously chosen parents. For each selection candidate, the distance to previously selected settings is calculated as follows: 

\[ d_{j} = \min_{\{k | (x_{k}) \text{ is a previously selected setting}\}} \sqrt{\sum_{i=1}^{p} \left(\frac{x_{i,j} - x_{i,k}}{\hat{\sigma_i}}\right)^2} \]

with:
\begin{itemize}
    \item[--] \( p \) is the number of parameters.
    \item[--] \( \hat{\sigma}_{i} \) denotes the empiricial standard deviation of the \(i\)-th parameter in the current iteration.
\end{itemize}

This condition ensures a candidate gets selected only if its minimum distance from the existing pool of settings is larger than a defined threshold. For step 3.1, the threshold is $0.04 \cdot n$, while given that step 3.2 is less affected by stochastic variations and more similar settings tend to have more similar values smoothed objective function, the threshold is raised to $0.08 \cdot n$. In contrast, for step 3.3, the sole constraint is that it has not been previously selected ($>0$). As the variance in individual parameters decreases throughout iterations, it implicitly leads to the selection of more similar settings in later stages and a more focused evolutionary process.

\subsection*{File S3}
To further improve convergence speed there are various minor improvements to consider regarding mutation rates. For class variables, in each iteration, the share of each class is calculated for both the pool of parametrizations and the selected parents. If for the last five iterations, the sum of the share of all but a single class is lower than the highest mutation rate (Step 4.3: $p_{activ,i} = 0.3$) and the share of these classes in the selected parameterizations in Step 4 is lower than $0.75 \cdot p_{activ,i} = 0.225$), mutation rates for this parameter are half. If conditions are still fulfilled for the reduced mutation rate, the mutation rate is further reduced to 0.01.

For continuous variables, additional prioritization and direction are given to some parameters instead of symmetrically sampling mutations around prioritizing one direction:
\begin{linenomath}
\begin{gather*} 
m_{size,i} \sim U(0, 2 \sigma_{x_i})\\
m_{dir,i} = 2 \cdot B(0, p_{sign,i}) - 1\\
m_{activ,i} = B(0, p_{activ,i})\\
t_i = m_{size,i} \cdot {m_{dir,i}} \cdot m_{activ,i}
\end{gather*} 
\end{linenomath}
with $p_{sign}$ on default being 0.5. To prioritize, the kernel regression is used to approximate the expected change in case of an increase or decrease in the parameter:
\begin{linenomath}
\begin{gather*} 
m_i^{+}(x) = \frac{m(x_1, \cdots, x_i + \hat{\sigma}_i, \cdots, x_p) - m(x_1, \cdots, x_i, \cdots,x_p)}{\hat{\sigma}_i}\\
m_i^{-}(x) = \frac{m(x_1, \cdots, x_i - \hat{\sigma}_i, \cdots, x_p) - m(x_1, \cdots, x_i, \cdots,x_p)}{\hat{\sigma}_i}\\
\end{gather*} 
\end{linenomath}
In case either of the two is positive, $p_{sign}$ is increased/decreased by 0.03 to make it more likely to sample in that respective direction. In case both $m_i^{'+}$ \& $m_i^{'+}$ are negative, the overall mutation probability $m_i$ is reduced by $20\%$. This procedure is repeated for the optima from the last five iterations.
\par
In case at least five continuous parameters are considered, the local derivates for all parameters in the current iteration are compared and the mutation rate $m_i$ in the 20\% of the parameters with the most favorable mutation rate are increased by $50\%$. Contrarily, the least favorable mutation rates are reduced by $50\%$. In case a mutation rate exceeds 0.3 / 0.4 for Step 4.2 / 4.3 it is reduced to this respective threshold.  
\par
To avoid non-impactful mutation in case a parameter is fixed, is also advised to specify a minimum value for the mutation range that is then used instead of $2 \cdot sigma_i$. For all scenarios in this study, a minimum range of 40, 20, and 3 have been used for $x_1$, $x_2$, and $x_3$, respectively.

\end{document}